# Integrating Structural Description of Data Format Information into Programming to Auto-generate File Reading Programs

**Xinghua Cheng[a, d], Erjie Hu[a, c], Di Hu[a, b, c *]**

[a] School of Geography, Nanjing Normal University, Nanjing, 210023, China;
[b] Jiangsu Center for Collaborative Innovation in Geographical Information Resource Development and Application, Nanjing, 210023, China;
[c] Key Laboratory of Virtual Geographic Environment (Nanjing Normal University), Ministry of Education, Nanjing, 210023, China;
[d] Department of Land Surveying and Geo-Informatics, The Hong Kong Polytechnic University, Kowloon, Hong Kong;

***Correspondence**: hud316@gmail.com; School of Geography, Nanjing Normal University, No.1 Wenyuan Road, Xianlin University District, Nanjing, China.

**Link to code:** https://github.com/flute0316/DFMLEditor

## Highlights

- Describing data format information of files by Data Format Markup Language (DFML)
- Automatic generation of codes for reading files
- Sequentially and randomly reading files

**Authorship Statement:** Xinghua Cheng and Di Hu designed the research flow and wrote the manuscript. Xinghua Cheng and Erjie Hu performed the experiments in the study. A-Xing Zhu and Guonian Lü contributed significantly to framing the scientific research question of the study and provided constructive discussion. All authors read and approved the final manuscript.

# Abstract

File reading is the basis for data sharing and scientific computing. However, manual programming for file reading is labour-intensive and time-consuming, as data formats are heterogeneous and complex. To address such an issue, this study proposes a novel approach for the automatic generation of file reading programs based on structured and self-described data format information. This approach provides two modes (i.e., sequentially and randomly reading) and is composed of three parts: (a) the file data format is described by Data Format Markup Language (DFML) and thus DFML documents are generated; (b) formation of data type sequences by parsing those DFML documents; (c) generation of programs for sequential (or random) reading data with formed data type sequences and general programing rules for specific programming languages. A tool named "DFML Editor" was developed for generating and editing DFML documents. Case studies on binary (ESRI point shapefiles) and plain text files (input files of Storm Water Management Model) were conducted with the software developed for automatic program generation and file reading. Experimental results show that the proposed approach is effective for automatically generating programs for reading files. The idea in this study is also helpful for automatically writing files.

## 1. Introduction

Data reading is the basis of data sharing and scientific computing. Since a large part of the data in these fields are stored in files, data reading usually involves file reading. Handling file reading for a large number of files in different data formats is a major headache in many fields, such as geoscience, biology, physics, and chemistry.

The approaches of addressing file reading problems can be generally classified into two kinds: (i) specific programming for reading files with specific data formats–one program for one data format–and (ii) universal programming for reading files with different kinds of data formats–one program for several different data formats. In the former, for a file in a specific data format, we need to make a specific program or package to read the file. For example, Buckley (2015) provided a pythonic package for reading files in the Supersymmetry Les Houches Accord format (Skands et al. 2004). Clearly, this kind of programming has many deficiencies. When equipment providers define their data format for storing data, the generation of a specific program for reading data in that data format requires the ability to understand the data format and to realize the syntax rules of programming languages (Ballagh et al. 2011).

In the latter approach, the data formats of files can be considered a general framework, e.g., Environment Systems Research Institute (ESRI) shapefiles (ESRI 1998) and Word and Excel files. These files are different in terms of data type, data structure, and data layout (Hu et al. 2015), though they belong to the same kind of data format. As one can imagine, reading these files is not an easy task since it depends on the complexity of the data structure, data layout, data type and so on. Currently, a feasible solution is to integrate all program modules that read files of various data structures into a package. Many efforts have been made by researchers in different fields, e.g., the ProteomeCommons.org IO Framework for reading and writing proteomics data with multiple formats (Falkner et al. 2006); the RNetCDF–A package for reading and writing NetCDF datasets (Michna and Woods, 2013); and Open Babel (O'Boyle et al. 2011), an open chemical toolbox integrating many cheminformatics algorithms for reading chemical files. It is worth noting that once changes are made to any part of a file, we should manually modify one of the program modules (sometimes even the whole program) to enable reading the data correctly. This means that this kind of programming approach relies heavily on manual maintenance.

From the content mentioned above, we can find that the generation of file reading programs is strongly related to the data formats. The diversity of data formats and the complexity (e.g., structure and layout) of files lead to challenges in generating programs for reading files. Manual programming is labor intensive and time consuming. To address these issues, this study proposes an approach based on structured and self-described data format information to automatically generate programs for reading files. The remainder of this article is organized as follows. Section 2 analyzes the data formats and the principles of file reading. Then, a strategy to automatically generate programs for reading files is proposed in Section3 followed by Section 4 which explains the strategy in detail, i.e., the formation of a linear data type sequence for a file and the generation of a program based on such a sequence. Experimental validations of the proposed approach using two cases are reported in Section 5. Finally, a discussion and some concluding remarks are presented in Section 6.

## 2. Analysis of data formats and file reading

### 2.1 Analysis of file data formats

The premise of reading a file is a detailed understanding of the formats of data stored in that file. The data format of a file in information technology refers to a format describing how data are organized and stored in a specific file system (Korn and Vo, 2002; Larobina and Murino, 2014). Programming for reading files heavily relies on an understanding of the data format. In other words, programmers need to prepare codes strictly in accordance with data formats. As shown in Fig 1, when we want to read a file and extract its contained content, we should first read its data format description document, thereby understanding its data storage rules.

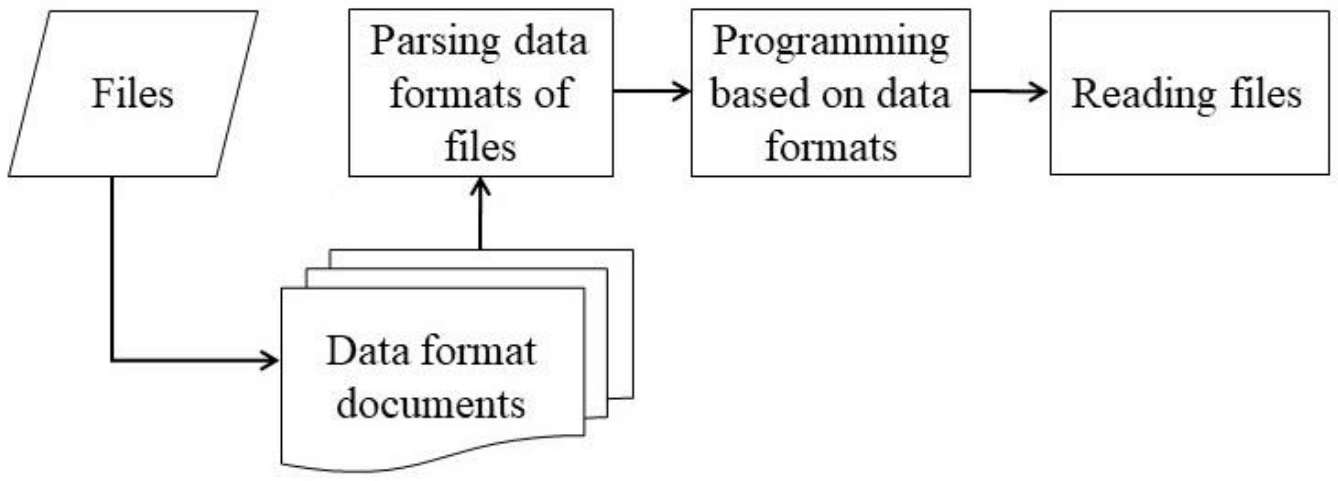

**Fig. 1.** Schematic expression of the relationship between data formats and programming.

Various data formats can be found in various fields, and the understanding of data formats is unclear and inconsistent. For example, a plain text file can be regarded as a general data framework. However, the content of a plain text file can be very different in terms of the data structure and data type. In this study, we hold the view that the variety of data formats is determined by three parts, including data type, structure, and layout. Furthermore, Fig 2 shows a conceptual model for describing a data format composed of four entities (i.e., data types, groups, separators, and locations). A group entity represents the structure of the data and is associated with three entities: data types, separators and locations. The distribution of group entities forms the layout of data. Specifically, a separator and location result in a different data structure and layout. According to the conceptual model, the data format is defined as the efficacious amalgamation of the data types, structure, and layout.

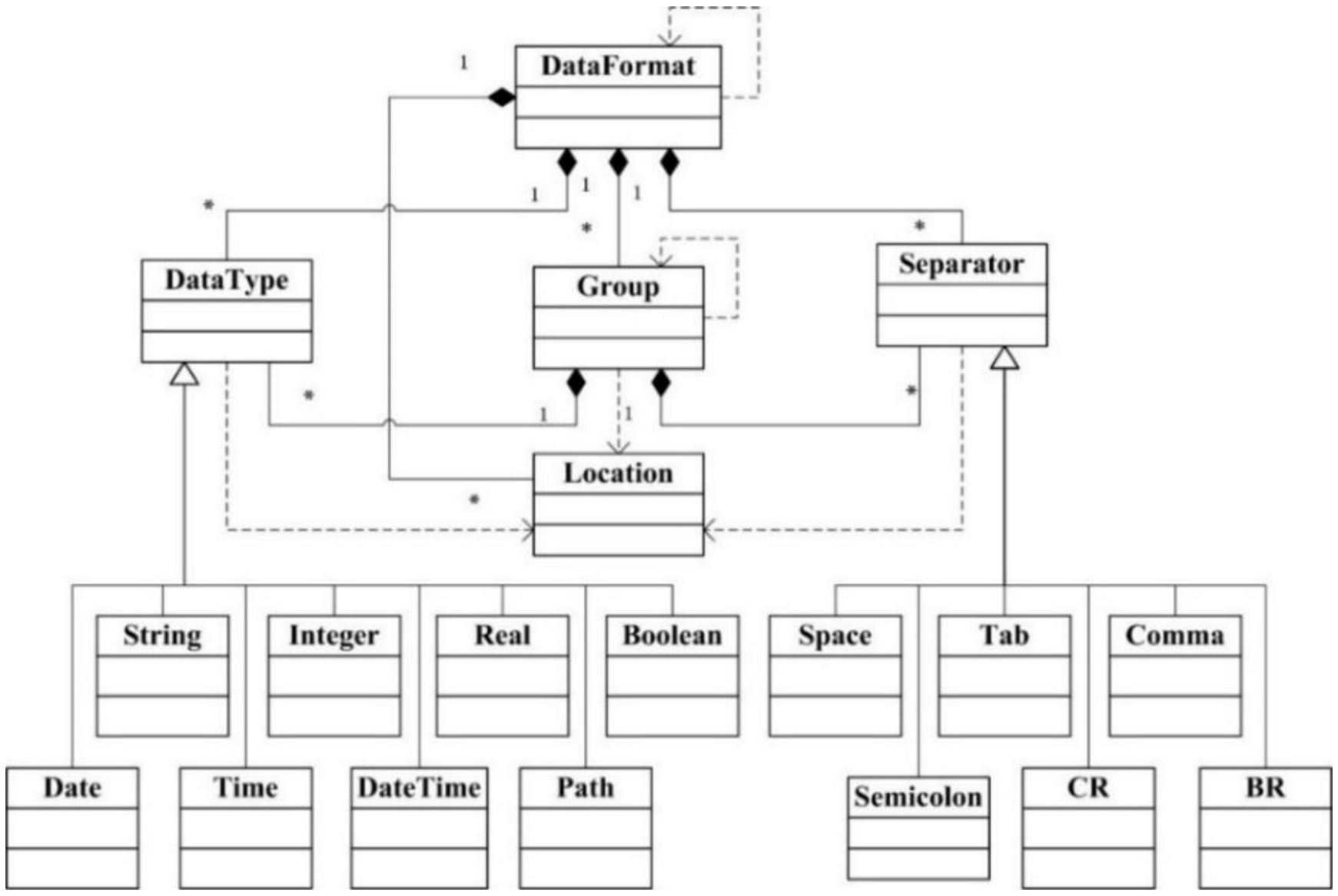

**Fig. 2.** A conceptual model of data formats (Hu et al. 2015).

## 2.2 Analysis of basic file reading principles

The principles that computers use to read a file can be described linearly. When we want to read a data item in a file, we need to make programs according to the byte location, byte length and data type. To describe file reading conveniently, a binary file consisting of five sections (i.e., S1, S2, S3, S4 and S5), as illustrated in Fig 3, is taken as an example. Each section includes different data items and is discerned by its start location and end location. For example, the start and end locations for S1 are 0 and 53, respectively. When the computer wants to read the data items *a*, *b*, and *c*, their locations, byte lengths and data types should be provided. Note that a data item may not appear only once in a file. For instance, data item *a* exists in both S1 and S5, appearing up to three times. At this point, we need to know the location, byte length, and data type of each data item *a* if we want to randomly read it.

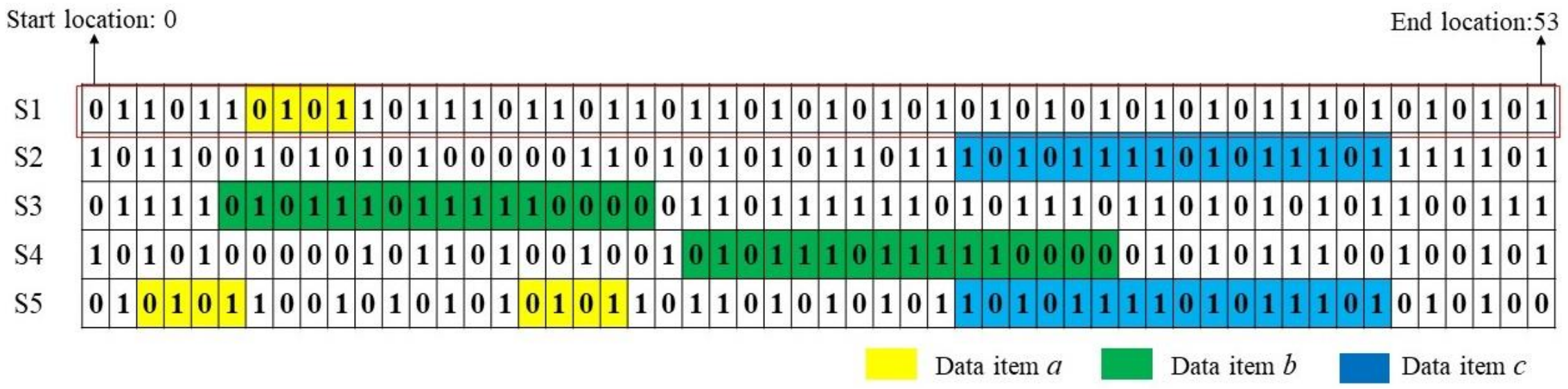

**Fig. 3.** An example of a binary file

In this study, the location of a data item can be considered to be absolute or relative. The absolute location represents the position in the whole file, whereas the relative location is the position in a specific file section. Fig 4 shows two methods, namely, absolute and relative, of describing the location representation for the same data items. The start location of data item *c* is 32 in S5 and is 249 in the whole file.

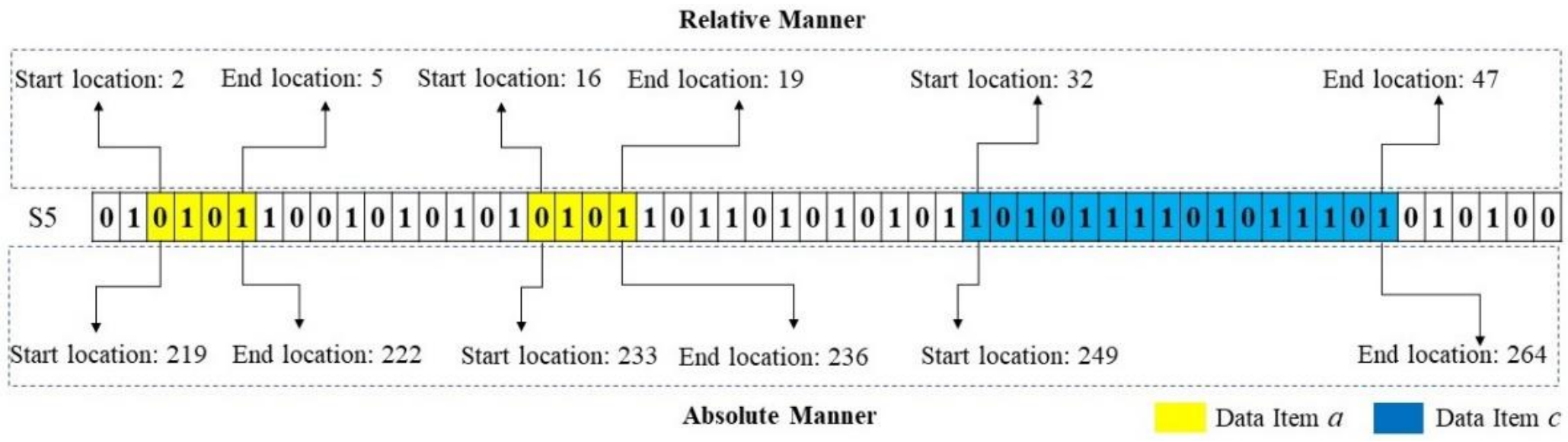

**Fig. 4.** An example of absolute and relative locations for S5.

# 3. Integrating structural data format information into programming: A strategy

## 3.1 General considerations for generating a file reading program

The analysis in Section 2 demonstrates that for automatic generation of a program for reading a file, three parts should be considered as follows:

(1) generation of the basic program framework

(2) generation of the function modules for reading data contained in the file

(3) reading methods (i.e., sequential reading and random reading)

The first part varies among programming languages since their syntaxes are different. Table 1 shows the code structure and components using C# (Microsoft 2017). The second part can be implemented by individually analyzing the data format of a file. The codes for reading each part of a file are made according to the data type, structures, and layouts of this part and are finally integrated into a program. The third part influences the generation of codes and is determined by the choices made by users. For sequential reading, we need to generate codes to read the whole file. In regard to random reading, the basic data format information for a specific data item should be declared clearly.

**Table 1** Code structure and its components in a C# program.

| Code structure | Components of code structure |
|---|---|
| using namespace | keywords of the namespace, name of the namespace |
| namespace definition | keywords of namespace definition, name of the namespace |
| type definition | access modifier, type keyword, type names |
| constructor definition | access modifier, name of constructor, field name, property type |
| field (or property) definition | access modifier, type of field (or property), name of the variable |
| method definition | access modifier, return type, method name, method parameter |

## 3.2 Structural description of data format information with DFML

The structural description of the data format of a file plays an essential role in generating programs in an automatic manner. In this study, Data Format Markup Language (DFML) (Hu et al. 2015) is employed to describe the data format of a file. Based on Extensible Markup Language (XML), DFML makes it possible to exactly describe various data formats in a universal manner. With the use of DFML, the structures and data types of items in a file can be efficiently described by a DFML document. Six XML elements including the data format element (root element), import element, location element, data type element, separator element and group element, are designed in DFML and tabulated in Table 2. Moreover, Fig 5 shows how to use DFML to describe the data format of a file.

**Table 2** Description of key XML elements in DFML (Hu et al. 2015).

| XML element | Attribute | Description |
|---|---|---|
| Data format | Name | The name of the described data format |
| | Mode | Represents text file format and binary file format using character mode and byte mode |
| Import | Link | The location of the imported data format markup document |
| Location | Name | The name of the location |
| | | A named location can be quoted in the context |
| | Value | The value of the location's starting position and ending position |
| Data type | Name | The name of a new data type, only used for expandability |
| | Type | The named data type can be quoted in the context |
| | | The specific data type of the data type element |
| | | The value range includes the 8 basic data types and other types for expandability |
| | Format | The presentation format of a special data type |
| | Number | The number of repeating data type elements |
| | Location | The location of a data type element |
| Separator | Name | The name of the separator element |
| | | A named separator element can be quoted in the context |
| | Type | The specific separator of the separator element. The value range includes the 6 basic separator types and other separators for expandability |
| | Number | The number of repeating separator elements |
| | Location | The location of a separator element |
| Group | Name | The name of the group element |
| | | A named group element can be quoted in the context |
| | Location | The location of the group element |
| | Number | The number of repeating group elements |

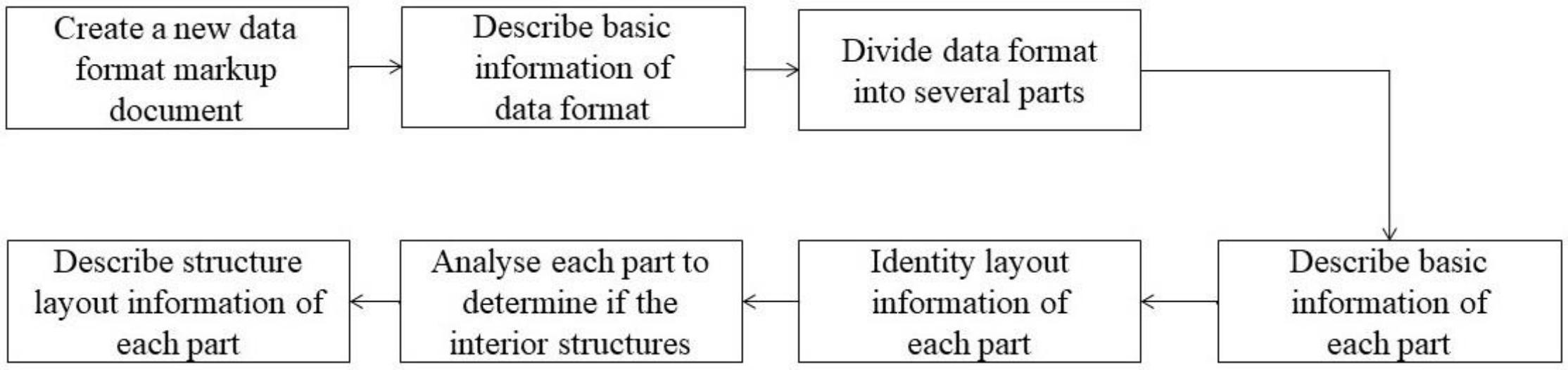

**Fig. 5.** Description of the workflow of the DFML data format (Hu et al. 2015).

## 3.3 A strategy for automatic generation of file reading programs

By integrating general programming considerations and the structural description of the data format information of a file, a strategy is proposed as follows:

(1) The data format information of a file is structuralized by DFML;

(2) A linear data type sequence is formed by parsing the DFML document;

(3) Automatic generation of a file reading program by parsing the formed linear data type sequence.

In this proposed strategy, data type sequence means the sequence composed of data items with different data types (e.g., integer, double and string). Based on the location representation of a data item in a file mentioned in section 2.2, the linear data type sequence falls into two categories: absolute and relative. The former is employed for random reading, whereas the latter is adopted to sequentially read data. A flow chart of the implementation based on the proposed strategy is described in Fig 6.

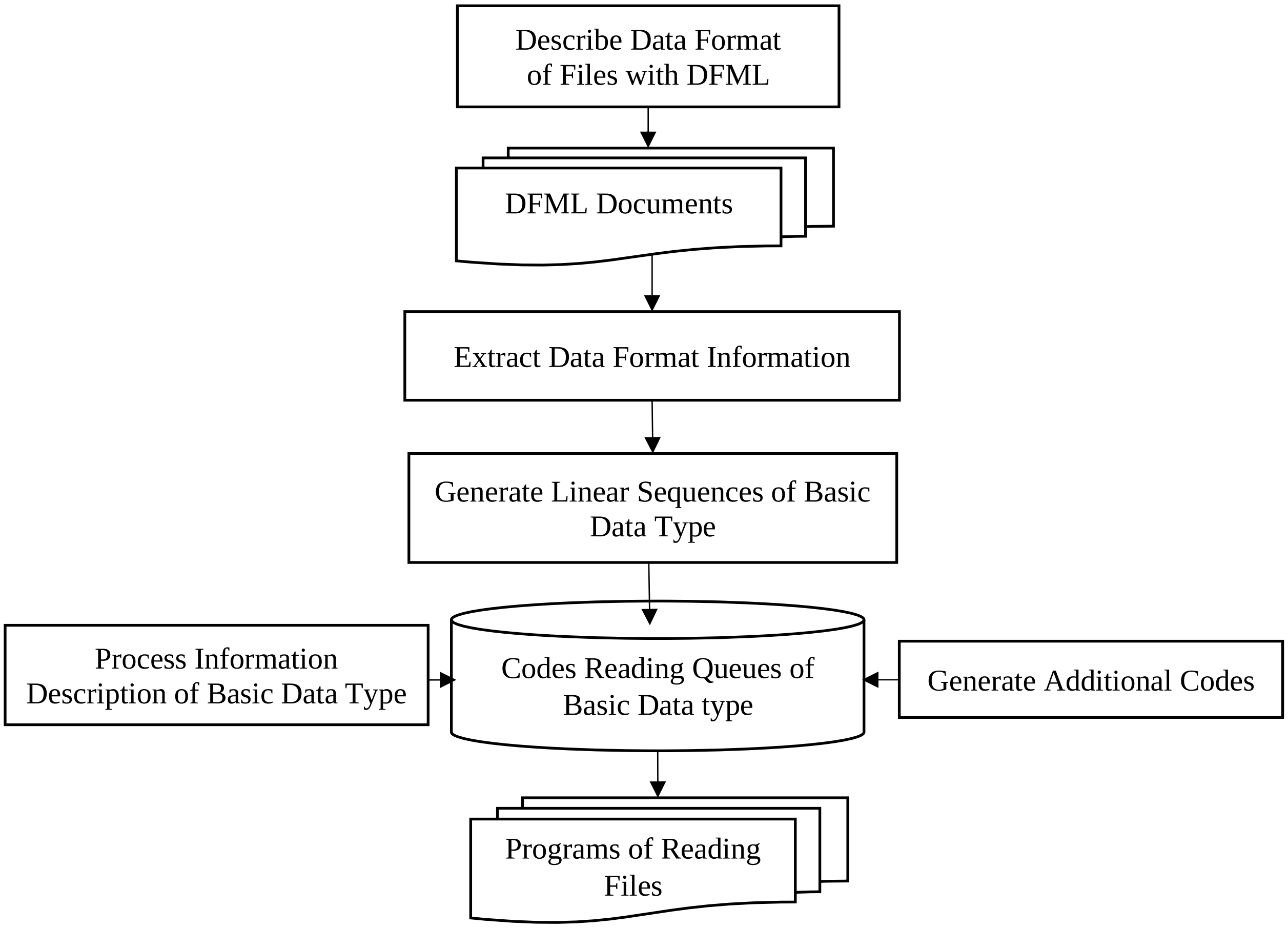

**Fig. 6.** Schematic diagram for the automatic generation of file reading programs.

# 4. Automatic generation of a file reading program

## 4.1 Forming data type sequences with structuralized data format description information

The data format information of a file is the key to generating codes for reading it. In this study, the data format description document for a file is generated by using DFML. Then, this document is converted to a data type sequence that is used to generate codes to sequentially or randomly read the data contained in the file. The data type sequence is indeed the ordered collection of XML elements. Each XML element records the format information including the data type, data location, default value, byte order, comments, occurrence times and data length interval before the next occurrence. The following steps show how to convert a DFML document to a data type sequence:

(1) Recognizing the data format description information in units of data types

The mark for a data type is taken as the identification symbol. The original properties of an individual XML element will be reserved. Additional properties, such as the occurrence times and data length interval, are added to the properties of the XML element for enhancing the efficiency of the following program generation. Concretely, the occurrence times of an individual data (item) type are the ratio of the data length of the whole file to that of itself. The data length interval is the length of other data between the location of a data item and that of its next occurrence.

(2) Recognizing the group structure and then converting data items in this group to a linear data type sequence

In response to the group mark of a group structure, the data format information of multiple data items contained in the group is successively extracted. The starting location value of the location property in the group structure is taken as the starting location of the first child data item. The remaining data items in this group are successively analyzed with location properties and the data lengths of a parent XML element and themselves and then individually converted to XML elements with additional properties. The repetitive occurrence times of a data item are calculated as the ratio of the length of the parent data item to its own data length. Specifically, when the length of a file or a group data item is unknown, the repetitive occurrence times of a child element can be assigned a value of “-1”.

(3) Generating the ordered XML element sequence in units of data type

The XML element sequence is the set of data item in units of data type extracted from the DFML document. In this set, the XML elements are ranked in ascending order of the start locations of their corresponding data items in a file.

To further describe these steps, pseudocode is shown in Fig 7.

```
rootElem: The root element of the DFML document
linearSequence: The object storing the XML elements
startLocation: The start location of a data item
endLocation: The end location of a data item

For all element in rootElem:
    elemName = Acquire the name value of element
    locationAtt = Acquire the "location" attribute value of element
    startLocation, endLocation = Parse the start and end location form locationAtt
    If elemName equals "group":
        interval = Sum (length of child elements included by element)
        If endLocation = "-1":
            groupLength = "-1"
            repetition = "-1"
        Else:
            groupLength = endLocation – startLocation
            repetition = The ratio of groupLength and interval
        End
        For all childElem in element:
            childLocationAtt = Acquire the "location" attribute value of childElem
            childStartLoc = Parse the start location from childLocationAtt
            childEndLoc = Parse the end location from childLocationAtt
            childStartReadLoc = startLocation + childStartLoc
            childLength = childEndLoc - childStartLoc
            Add childStartReadLoc to attributes of childElem as "startReadLocation"
            Add childLength to attributes of childElem as "length"
            Add interval to attributes of childElem as "interval"
            Add repetition to attributes of childElem as "repetition"
            Add childElem to linearSequence
        End
    Else:
        interval = 0
        repetition = 1
        elemLength = endLocation – startLocation
        Add startLocation to attributes of element as "startReadLocation"
        Add elemLength to attributes of element as "length"
        Add interval to attributes of element as "interval"
        Add repetition to attributes of element as "repetition"
        Add element to linearSequence
    End
End
Return linearSequence
```

**Fig. 7** Pseudocode for converting a DFML document to a linear data type sequence.

## 4.2 Generating a program with the formed data type sequences

The formed XML element data type sequence records the detailed information of data items included in a file. At this point, we just need to read each XML element and then make corresponding codes based on its properties such as data types, location, byte order, data length and data length interval. The steps of the automatic generation of a program can be described as follows:

(1) Generating codes of reading the data in a file with data type and location information in each XML element

According to the location properties of an XML element, the file pointer is positioned at the beginning location of a file. The corresponding data content is read by the file pointer in accordance with the data length, repetitive occurrence times, and data length interval. Two modes of generating codes can be used: sequential reading and random reading. Regarding the latter, the codes are generated with respect to the data items selected by users.

(2) Unifying the format and byte order of reading data and then generating corresponding codes

Concerning a binary data file, the storage method of a data item is examined. For example, if the byte order of a data item is large, then a specific method is called to convert the large byte order to a small one.

(3) Generating codes for parsing a data item and acquiring the corresponding data value

According to the data type of a data item, codes are generated for reading the corresponding data. In response to reading a data item that appears many times in a file, repeat steps (1) to (3), integrate all codes together, and save them as a file. Pseudocode for implementing these steps is shown in Fig 9.

```
linearSequence: The object storing the XML elements

For all element in linearSequence:
    elemName = Acquire the "name" value of element
    startReadLocation = Acquire the "startReadLocation" attribute value of element
    readLength = Acquire the "length" attribute value of element
    interval = Acquire the "interval" attribute value of element
    repetition = Acquire the "repetition" attribute value of element
    Generating codes for startReadLocation and readLength
    If repetition equals -1:
        Generate codes to construct a while loop where the repetition times are
        unknown
    Else:
        Generate codes to construct a while loop where the repetition times are
        known
    End
    Generate codes to read data with readLength length from the startReadLocation
    Generate codes to add the read data to the list for storing the read data
    Generate codes to increase the value of startReadLocation by interval
    Generated code to reduce the value of repetition by one
    Generate codes to end a While loop
End
```

**Fig. 8** Pseudocode for generating codes based on a linear data sequence.

## 4.3 Implementation

As mentioned in section 3.3, describing data format of a file by DFML is the first essential step. In doing so, a DFML editor has been developed for generating and editing DFML documents. As shown in Fig 10, the DFML editor provides modules for generating and editing files. DFML elements are displayed as a tree. Figure 9 (a) shows the user interface where users can edit and add DFML elements. In Figure 9 (b), a DFML document is previewed.

(a)

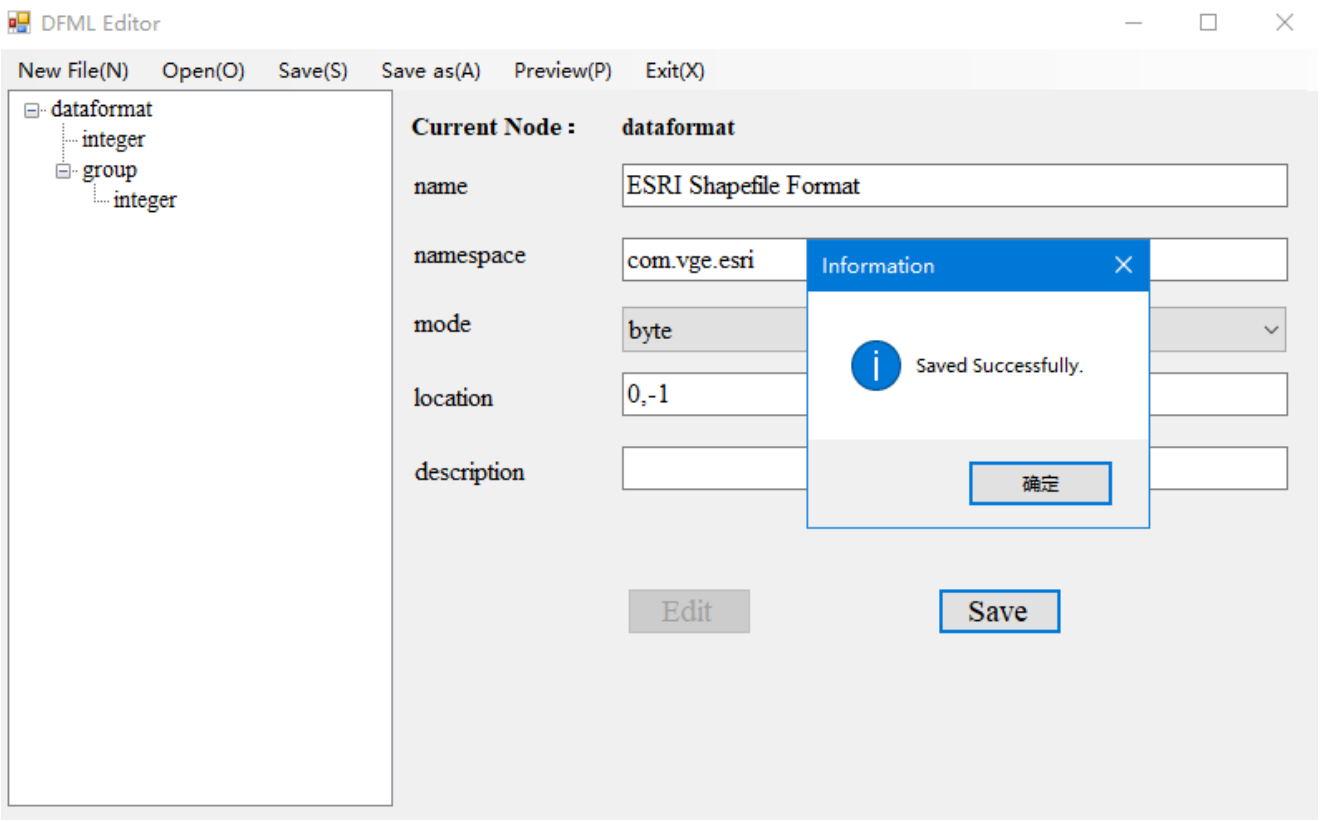

(b)

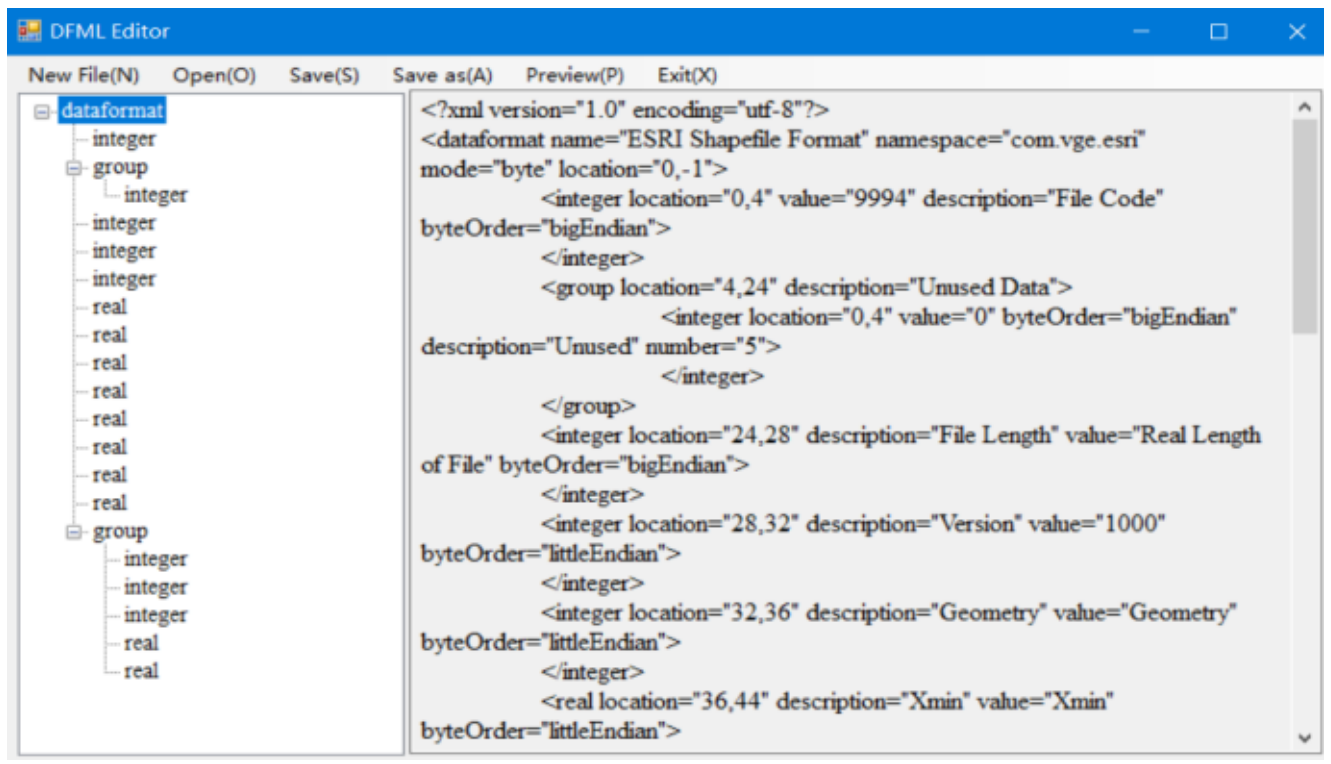

**Fig. 9**. DFML Editor. (a) Interface of the DFML editor, (b) Preview of the edited DFML document.

Fig 10 shows the implementation of the proposed strategy. There are three stages, i.e., loading the DFML document, generation of codes based on the inputted DFML document, and generation of outputs based on the codes. In the second stage, the DFML document is loaded and shown as a DFML tree to help us select the specific data items. Moreover, two modes (i.e., “sequentially reading” and “randomly reading”) are available for generating codes to read data. The former means the generation of codes for reading the whole file, whereas the latter is used to randomly read data.

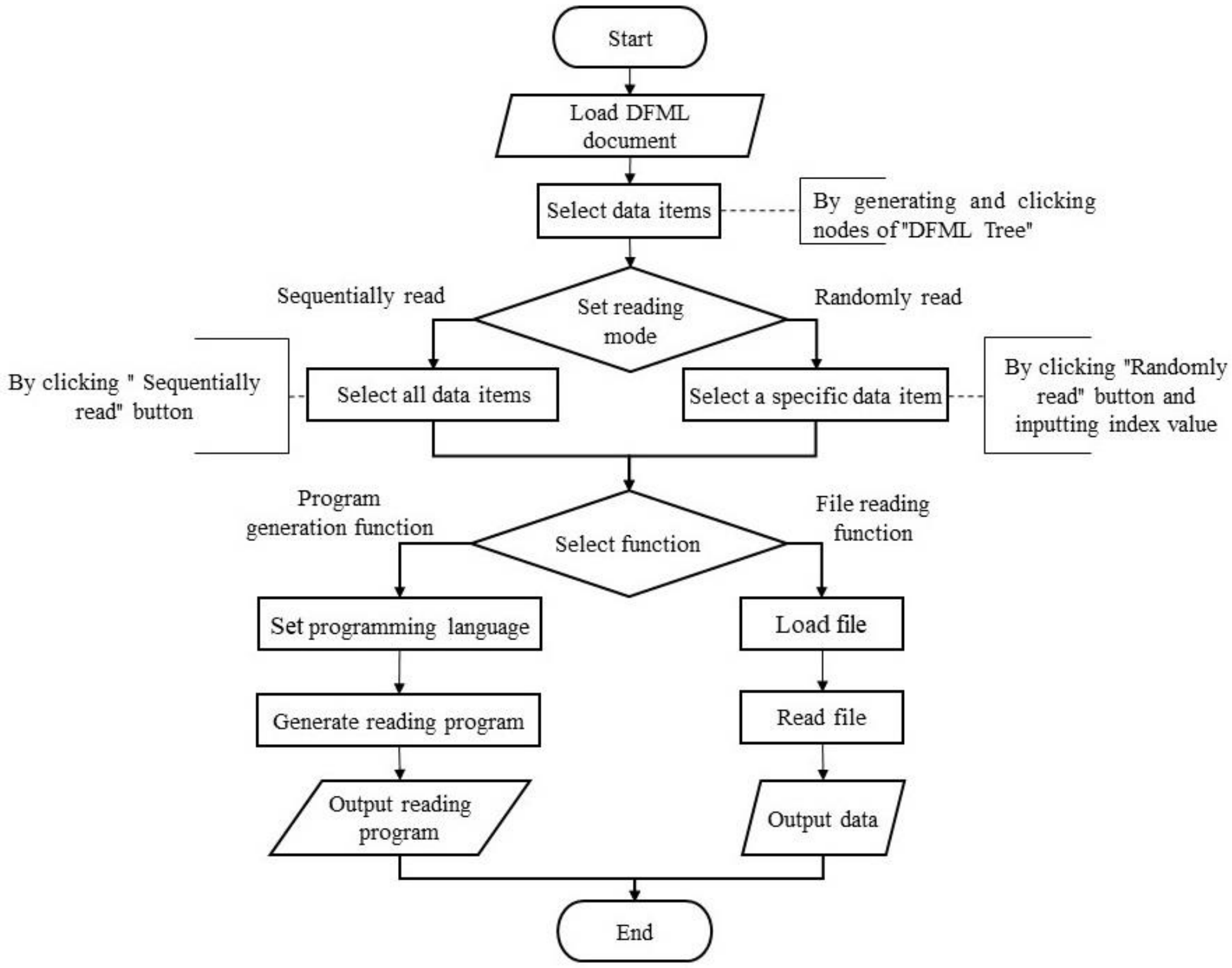

**Fig. 10**. Flowchart of generating codes and reading data by the proposed strategy.

Software has been developed to implement these stages. The programming language is C#, and the development environment is Microsoft Visual Studio 2019 with .NET Framework 4.7.2. The user interface of this software is shown in Fig 11.

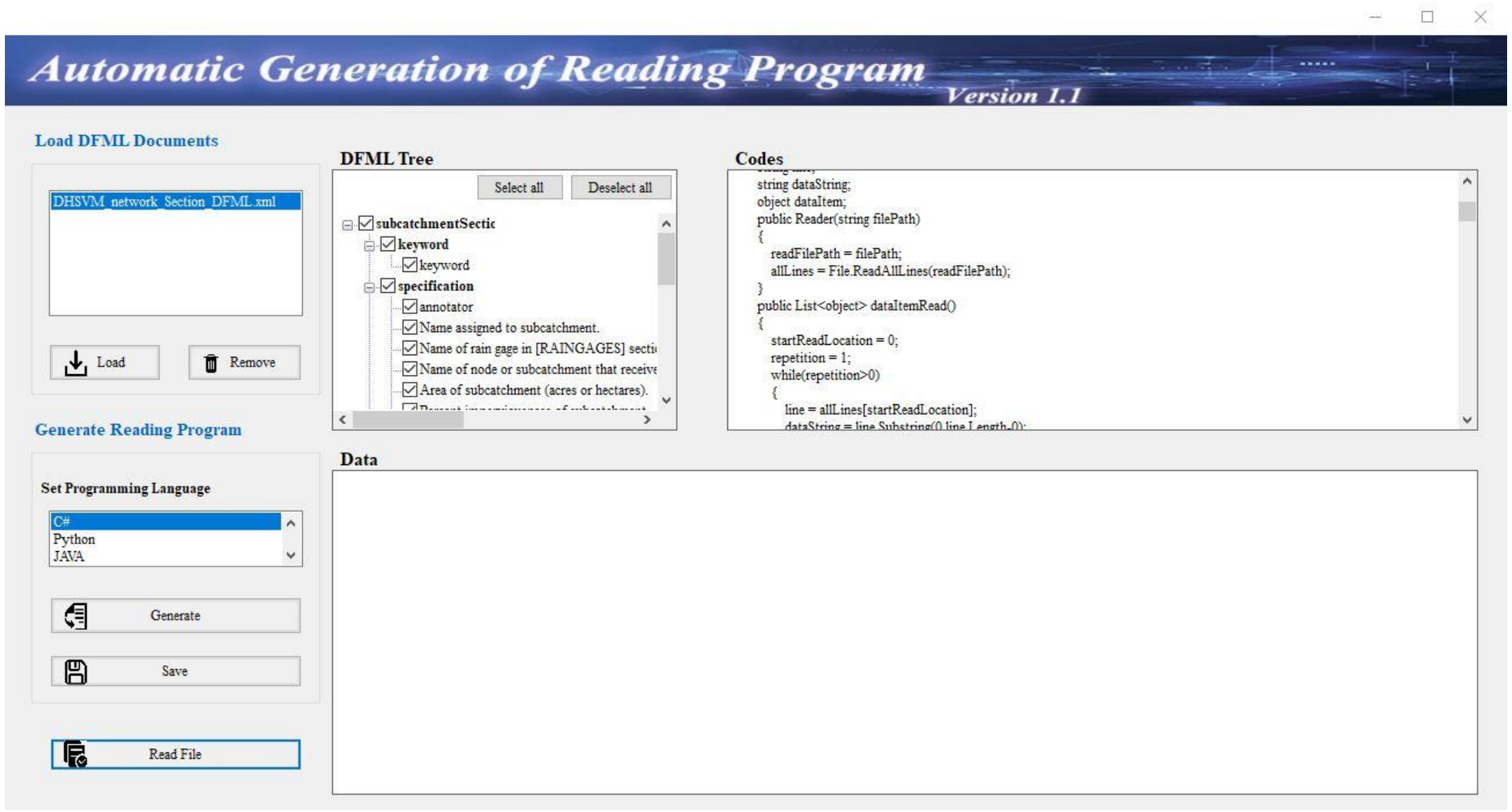

**Fig. 11**. The user interface of the developed software.

# 5. Case studies

## 5.1 Case study on binary files

This section describes reading a binary file, i.e., an ESRI point shapefile, which is widely used in the fields of surveying, mapping and geo-information. This type of file consists of a main file (.shp), an index file (.shx), and a dBAse table (.dbf). The main file is used here. More specifically, it is composed of two parts, i.e., the file header and variable-length record contents. Fig 12 shows its organization. The main file header is 100 bytes long. Tables 2, 3, 4 and 5 show the details of the data format. The fields in the main header are described with position (the starting byte), value, type, and byte order.

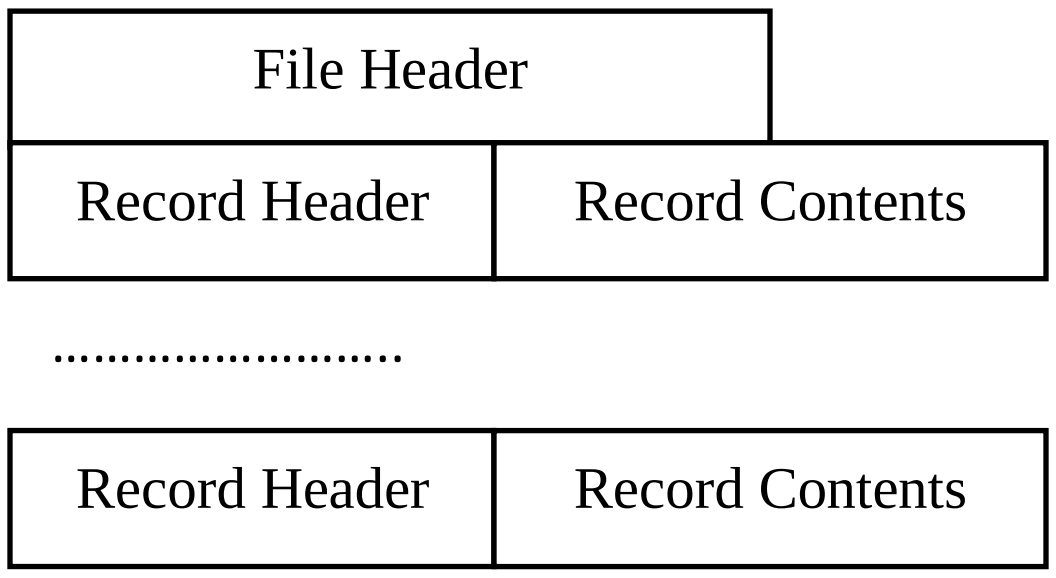

**Fig. 12** Organization of the main file.

**Table 2** Description of the main file header

| Position | Field | Value | Type | Byte Order |
|---|---|---|---|---|
| Byte 0 | File Code | 9994 | Integer | Large |
| Byte 4 | Unused | 0 | Integer | Large |
| Byte 8 | Unused | 0 | Integer | Large |
| Byte 12 | Unused | 0 | Integer | Large |
| Byte 16 | Unused | 0 | Integer | Large |
| Byte 20 | Unused | 0 | Integer | Large |
| Byte 24 | File Length | File Length | Integer | Large |
| Byte 28 | Version | 1000 | Integer | Large |
| Byte 32 | Shape Type | Shape Type | Integer | Small |
| Byte 36 | Bounding Box | Xmin | Double | Small |
| Byte 44 | Bounding Box | Ymin | Double | Small |
| Byte 52 | Bounding Box | Xmax | Double | Small |
| Byte 60 | Bounding Box | Ymax | Double | Small |
| Byte 68* | Bounding Box | Zmin | Double | Small |
| Byte 76* | Bounding Box | Zmax | Double | Small |
| Byte 84* | Bounding Box | Mmin | Double | Small |
| Byte 92* | Bounding Box | Mmax | Double | Small |

Note: * unused, with value 0.0, if not measured or Z type

**Table 3** Description of the record header of graphic format

| Position | Meaning | Value | Type | Byte order |
|---|---|---|---|---|
| Byte 0 | Graphic Type | Record Numbering | Integer | Large |
| Byte 4 | Record Length | Record Length | Integer | Large |

**Table 4** Description of the variable-length record of graphic format

| Position | Meaning | Value | Type | Byte order |
|---|---|---|---|---|
| Byte 0 | Graphic Type | Graphic Type (see in Table 4) | Integer | Small |
| Byte 4 | - | - | - | - |

**Table 5** Description of variable-length record contents of graphic format (point file)

| Position | Meaning | Value | Type | Number | Byte order |
|---|---|---|---|---|---|
| Byte 0 | Graphic Type | 1 | Integer | 1 | Small |
| Byte 4 | X | X | Double | 1 | Small |
| Byte 12 | Y | Y | Double | 1 | Small |

The DFML document for an ESRI point shapefile is shown as follows:

```
<dataformat name="ESRI Shapefile Format" namespace="com.vge.esri" mode="byte"
description ="location is with respect to byte 0">
```

```
<integer location="0,3" value = "9994" description = "File Code" byteOrder
="bigEndian" ></integer>
<group location="4,23">
   <integer value = "0" byteOrder="bigEndian" description="Unused" number="5">
   </integer>
</group>
  <integer location = "24,27" description = "File Length" value = "Real Length of File"
byteOrder = "bigEndian"> </integer>
  <integer location = "28,31" description = "Version" value = "1000" byteOrder =
"littleEndian"> </integer>
  <integer location = "32,35" description = "Geometry" value = "Geometry" byteOrder =
"littleEndian"> </integer>
  <double location = "36,43" description = "Xmin" value = "Xmin" byteOrder = "littleEndian">
</double>
  <double location = "44,51" description = "Ymin" value = "Ymin" byteOrder =
"littleEndian"></double>
  <double location = "52,59" description = "Xmax" value = "Xmax" byteOrder =
"littleEndian"></double>
  <double location = "60,67" description = "Ymax" value = "Ymax" byteOrder =
"littleEndian"></double>
  <double location = "68,75" description = "Zmin" value = "Zmin" byteOrder =
"littleEndian"></double>
  <double location = "76,83" description = "Zmax" value = "Zmax" byteOrder =
"littleEndian"></double>
  <double location = "84,91" description = "Mmin" value = "Mmin" byteOrder =
"littleEndian"></double>
  <double location = "92,99" description = "Mmax" value = "Mmax" byteOrder =
"littleEndian"></double>
 <group location = "100, -1" description = "Point">
     <integer location = "0, 4" description = "Record Number" byteOrder =
"bigEndian"></integer>
     <integer location = "4, 8" description = "Content Length" byteOrder =
"bigEndian"></integer>
     <integer location = "8, 12" description = "Geometry Type" byteOrder =
"littleEndian"></integer>
     <real location = "12, 20" description = "X" byteOrder = "littleEndian"></double>
     < real location = "20, 28" description = "Y" byteOrder = "littleEndian"></double>
     </group>
</dataformat>
```

As shown in Fig 13, the DFML document has been loaded and shown as a DFML tree. The C# language has been used to generate codes. Moreover, all data items included in the shapefile are selected to be read. It can be easily seen that the data have been output in the user interface of the software. The corresponding values of the properties of point layer of 13 middle schools of Nanjing city, China, in 2008, are output. In regard to random reading, the codes for reading data are generated based on the selected nodes in the "DFML tree”. Fig 14 shows the codes for reading all properties of the third point. And its property values are also output.

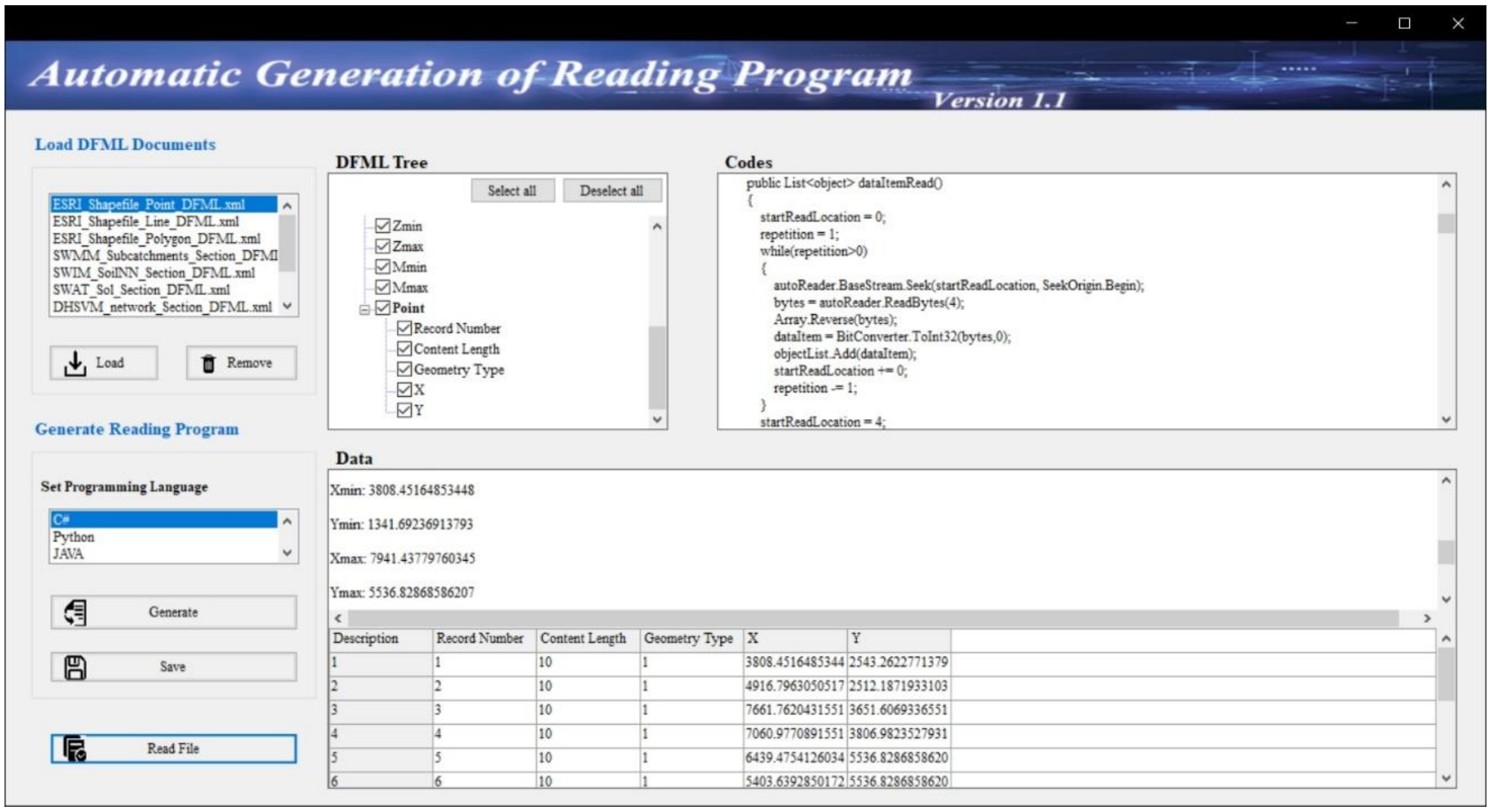

**Fig. 13** Automatic generation of codes for sequentially reading an ESRI point shapefile.

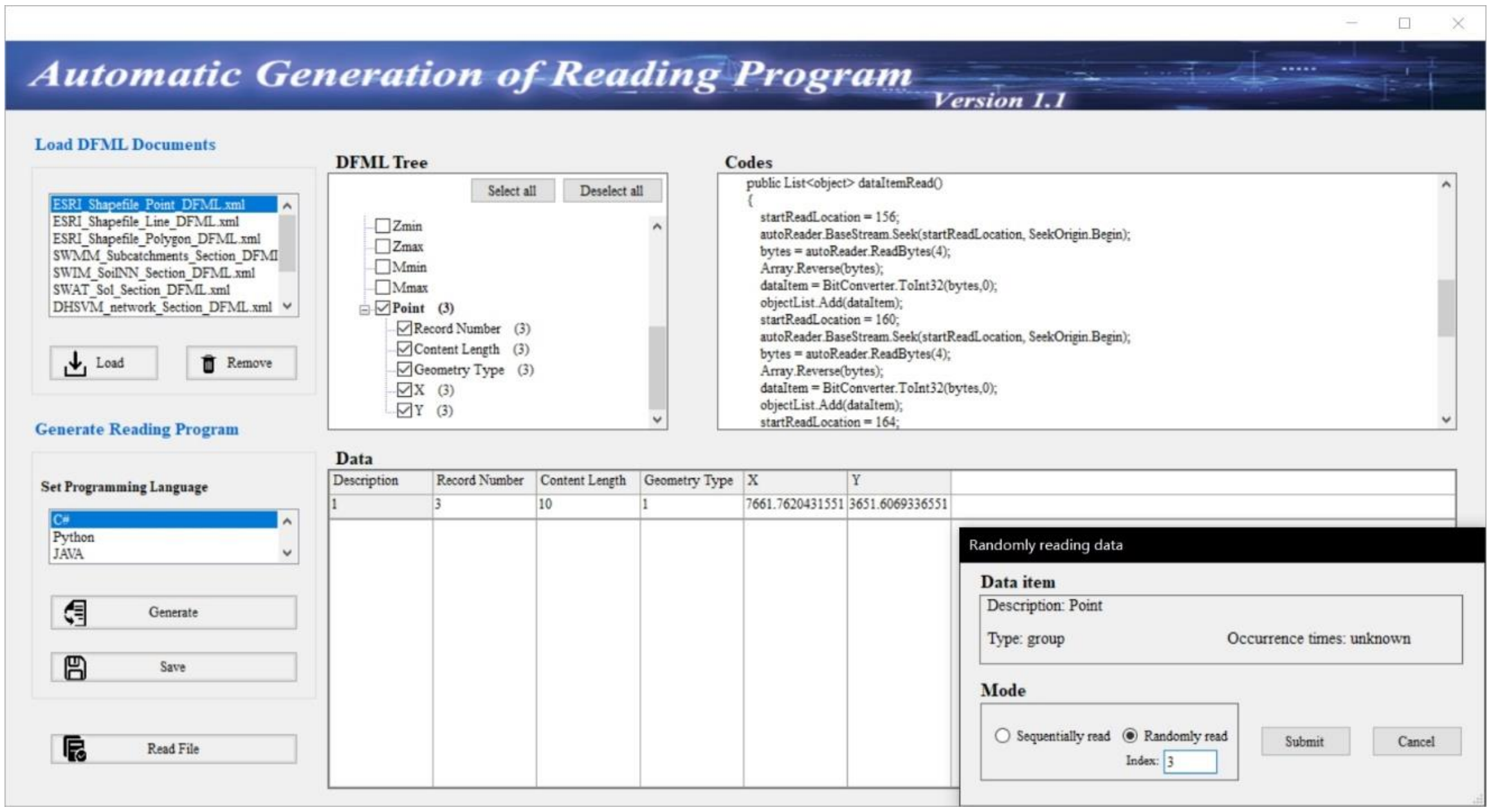

**Fig. 14** Automatic generation of codes for randomly reading data from an ESRI point shapefile.

## 5.2 Case study on plain text files

The second case study is reading a plain text file: one part of the input files of the Storm Water Management Model (SWMM), which is a dynamic rainfall-runoff simulation model, has been widely applied in hydrological studies (Burszta & Mrowiec 2013, Temprano et al. 2006). One part of the input data of the SWMM is stored in a .inp file consisting of 49 sections. A detailed description of such a file can be found in Rossman

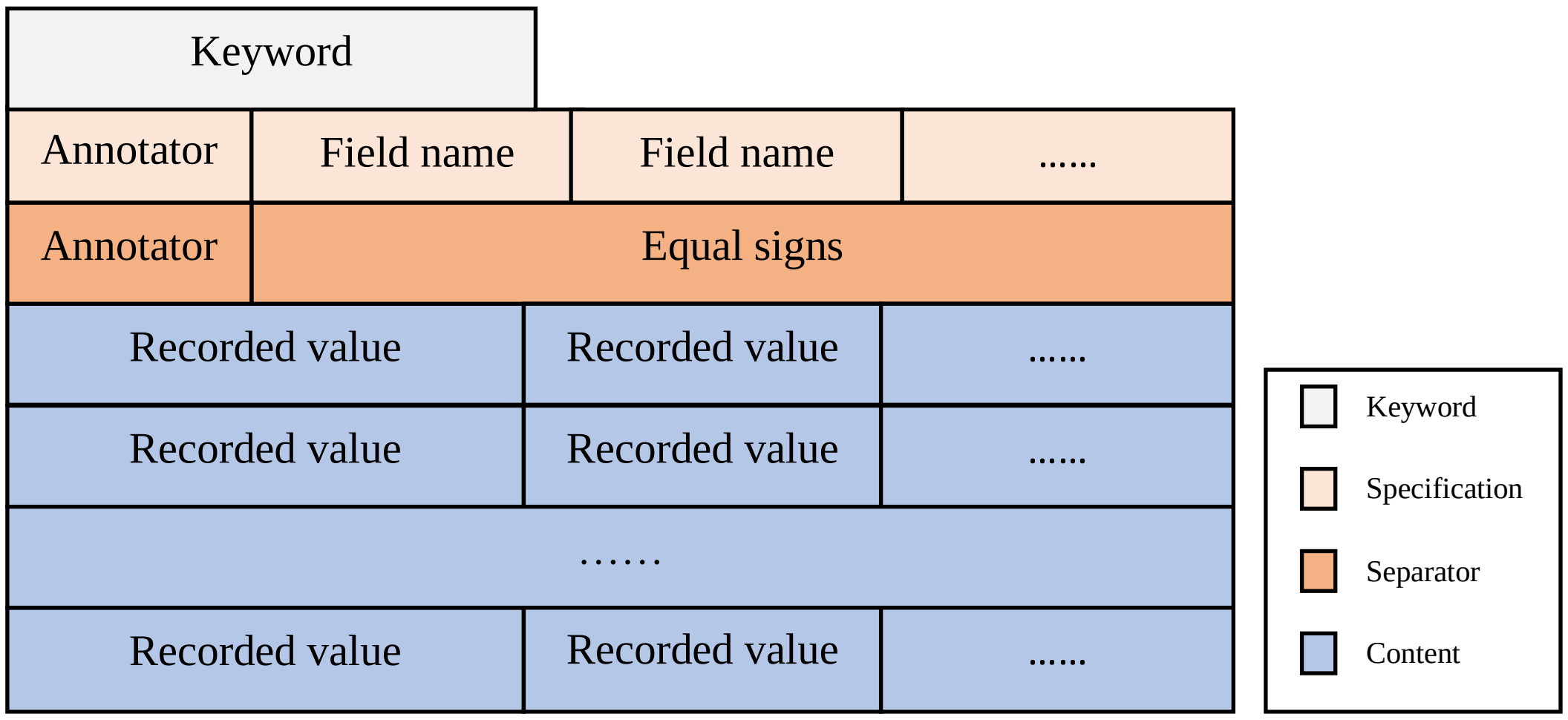

(2010). In this study, one typical section called *subcatchments* is used. More concretely, this section can be divided into keyword, specification separator and content parts. Fig 15 illustrates the organization of the *subcatchments* section.

**Fig. 15** Organization of the *subcatchments* section.

The keyword part is enclosed in brackets and indicates the type of this section. For example, the keyword part of the *subcatchments* section is "[SUBCATCHMENTS]". The specification part follows the keyword part, where double semicolons appear at the first two spaces as an annotator. The specification part is divided into several fields by blank spaces, and the fields vary in the different sections. Table 6 lists the fields for the *subcatchments* section.

**Table 6** Description of the fields in the *subcatchments* section

| Field | Position | Definition |
|---|---|---|
| Annotator | Space 0-1 | Double semicolons indicate the following is specification |
| Name | Space 2-6 | Name assigned to subcatchment |
| Rgage | Space 10-18 | Name of rain gage in [RAINGAGES] section assigned to subcatchments |
| OutID | Space 23-29 | Name of node or subcatchments that receives runoff from subcatchments |
| Area | Space 34-38 | Area of subcatchments (acres or hectares) |
| %Imperv | Space 43-50 | Percentage imperviousness of subcatchments |
| Width | Space 55-60 | Characteristic width of subcatchments (feet or meters) |
| Slope | Space 65-70 | subcatchments slope (percentage) |
| Clength | Space 75-82 | Total curb length (any length units) |
| Spack | Space 87-92 | Name of snowpack object (from snowpacks section) that characterizes snow accumulation and melting over the subcatchments |

The separator part consists of an annotator and following equal signs. This part does not store any data but is used only for separating the specification and content parts. A description of the separator part is given in Table 7.

**Table 7** Description of the separator part

| Field | Position | Value |
|---|---|---|
| Annotator | Space 0-1 | ; |
| Equal signs | Space 2-92 | = |

The content part consists of lines. Each line represents one record and can be divided into several record values by blank spaces. A description of one line of the content part is shown in Table 8.

**Table 8** Description of the content part in the *subcatchments* section

| Field | Position | Data type |
|---|---|---|
| Name | Space 0-10 | String |
| Rgage | Space 10-23 | String |
| OutID | Space 23-34 | String |
| Area | Space 34-43 | Real |
| %Imperv | Space 43-55 | Real |
| Width | Space 55-65 | Real |
| Slope | Space 65-75 | Real |
| Clength | Space 75-87 | Real |
| Spack | Space 87-92 | String |

Based on Tables 6, 7 and 8, a DFML document for the *subcatchments* section is generated and shown below.

```
<dataformat name="subcatchmentSection" namespace="com.vge.swmm.input" mode="char">
   <string description="section name" value="[SUBCATCHMENTS]" location="1 0,1 -
1"></string>
   <group description="section body" location="2 0,2 -1">
      <group description="table header" location="2 0,2 -1">
         <string description="annotator" value=";;" location="2 0,2 1"></string>
         <string description="Name assigned to subcatchment." value="Name" location="2 2,2
6"></string>
         <space location="2 6,2 10"></space>
         <string description="Name of rain gage in [RAINGAGES] section assigned to
subcatchment." value="Rgage" location="2 10,2 18"></string>
         <space location="2 18,2 23"></space>
         <string description="Name of node or subcatchment that receives runoff from
subcatchment." value="OutID" location="2 23,2 29"></string>
         <space location="2 29,2 34"></space>
         <string description="Area of subcatchment (acres or hectares)." value="Area"
location="2 34,2 38"></string>
         <space location="2 38,2 43"></space>
         <string description="Percent imperviousness of subcatchment." value="%Imperv"
location="2 43,2 50"></string>
         <space location="2 50,2 55"></space>
         <string description="Characteristic width of subcatchment (ft or meters)"
value="Width" location="2 55,2 60"></string>
         <space location="2 60,2 65"></space>
         <string description="Subcatchment slope (percent)." value="Slope" location="2 65,2
70"></string>
         <space location="2 70,2 75"></space>
         <string description="Total curb length (any length units)" value="Clength" location="2
75,2 82"></string>
         <space location="2 82,2 87"></space>
```

```
            <string description="Name of snowpack object (from [SNOWPACKS] section) that
characterizes snow accumulation and melting over the subcatchment." value="Spack"
location="2 87,2 92"></string>
            <space location="2 92,2 -1"></space>
            <cr></cr>
        </group>
        <group description="table separator" location="3 0,3 -1">
            <string description="annotator" value=";;" location="3 0,3 1"></string>
            <string description="separator" value="=" location="3 1,3 -1"></string>
            <cr></cr>
        </group>
        <group description="table content" number="unknown">
            <string description="Name" location="0 0,0 10"></string>
            <string description="Rgage" location="0 10,0 23"></string>
            <string description="OutID" location="0 34,0 43"></string>
            <real description="Area" location="0 34,0 43"></real>
            <real description="%Imperv" location="0 43,0 55"></real>
            <real description="Width" location="0 55,0 65"></real>
            <real description="Slope" location="0 65,0 75"></real>
            <real description="Clength" location="0 75,0 87"></real>
            <string description="Spack" location="0 87,0 -1"></string>
            <cr></cr>
        </group>
    </group>
</dataformat>
```

As shown in Fig 16, a DFML document is loaded for *subcatchments*. Codes for reading data are generated, and then all data of the *subcatchments* section are read and output. Regarding the “random reading” mode, Fig 17 illustrates examples of reading one specified data item.

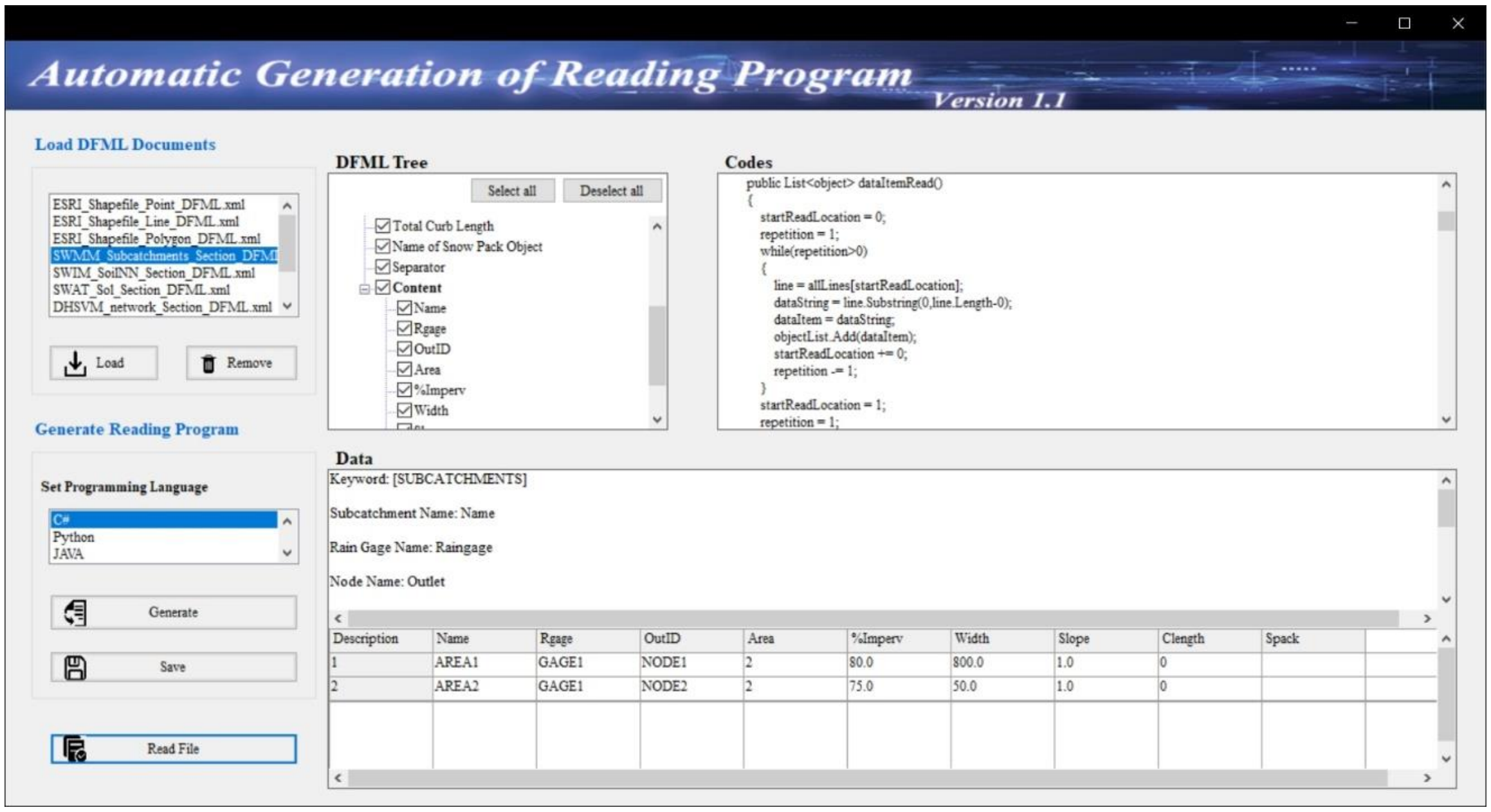

**Fig. 16** An example of sequentially reading a plain text file.

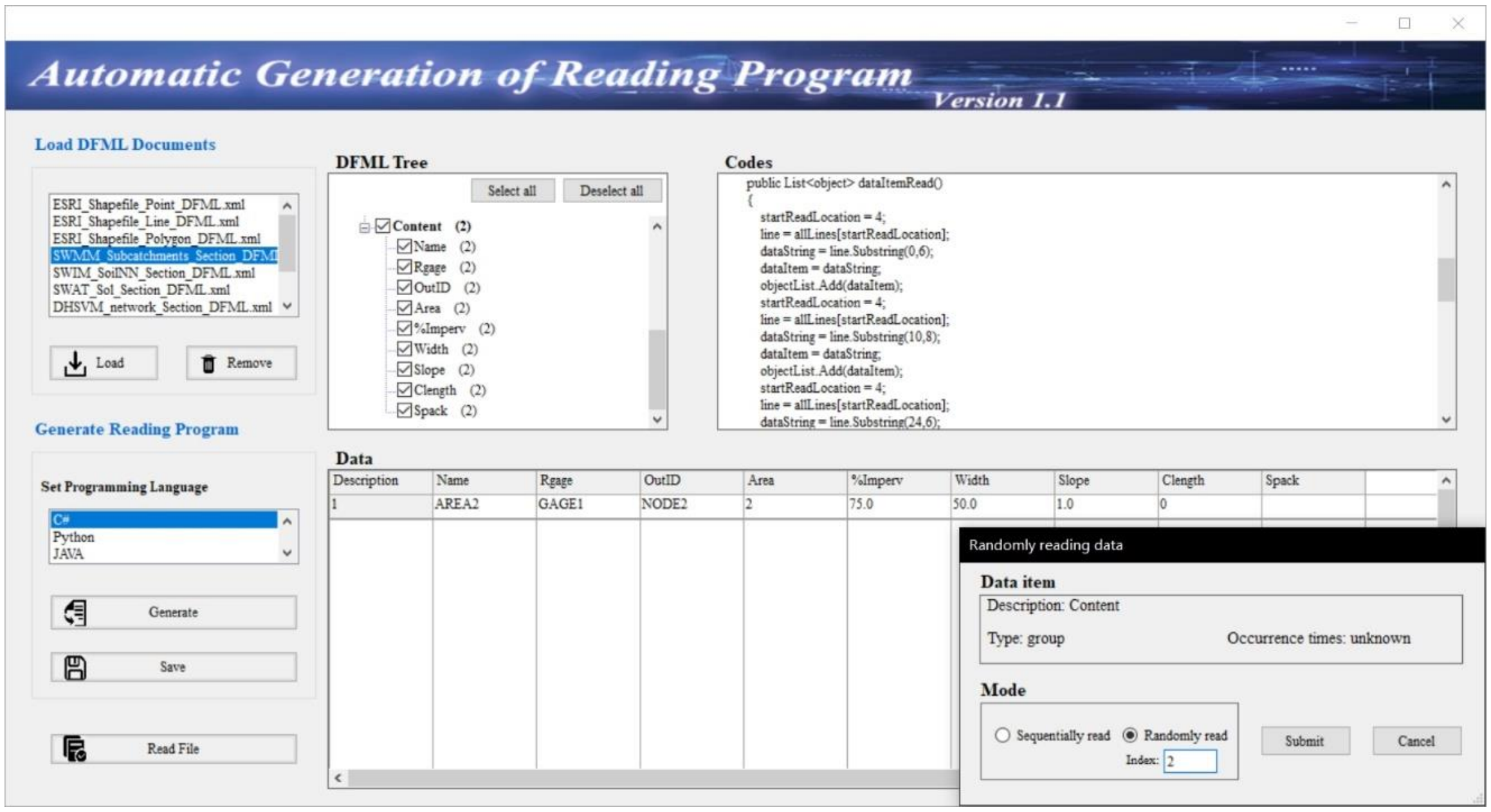

**Fig. 17** An example of randomly reading a plain text file.

# 6. Conclusions and discussion

File reading is essential for data sharing and scientific computing. Manual programming for reading files in heterogeneous data formats is labor intensive and requires the expertise of understanding data formats and different programming languages. To address such an issue, this paper proposes a novel approach for automatically generating a file reading program with the integration of the structural data format description. In this proposed approach, DFML is used to describe the data format of a file; thus, a DFML document is generated. Thereafter, a file reading program is automatically generated based on the data type sequence extracted from the DFML document. Two case studies were conducted on binary and plain text files. Experiments showed that the proposed approach is effective. The proposed approach is dedicated to enhancing scientific computing. Moreover, it is useful for constructing geographical models.

In this study, two aspects deserve further discussion. First , multiple group structures should be considered when a DFML document is converted into a linear data type sequence. A possible solution is to directly generate codes in accordance with the DFML document. Second, the methods of generating codes for reading data can be improved to generate more compact codes.

## Acknowledgments

**Funding**: This research is financially supported by the National Natural Science Foundation of China (Nos. 41771421, 42071365); National Key Research and Development Program of China (No. 2017YFB0503500); and Priority Academic Program Development of Jiangsu Higher Education Institutions (No. 164320H116).

**Conflicts of Interest:** The authors declare that they have no conflicts of interest.

## Appendix A.

Supplementary data to this article can be found online https://github.com/flute0316/DFMLEditor

## Computer Code Availability

Source code and software can be downloaded from the public GitHub repository called "DFMLEditor" (https://github.com/flute0316/DFMLEditor).

**Program language** C#; **Software**: .NET Framework 4.7.2

**Developers:** Erjie Hu, Xinghua Cheng, Di Hu.

**Contact address:** School of Geography, Nanjing Normal University, No.1 Wenyuan Road, Xianlin University District Nanjing, China.

**Year first available:** 2020

**The contact telephone number:** (+86) 152-5099-2342

**E-mail:** hud316@gmail.com

**Program size**: about 784 KB.